\documentclass[12pt]{article}
\usepackage{fancyhdr}
\usepackage{extramarks}
\usepackage{amsmath}
\usepackage{amsthm}
\usepackage{amsfonts}
\usepackage{siunitx}
\usepackage{tikz}
\usepackage[plain]{algorithm}
\usepackage{algpseudocode}
\usepackage{multirow}
\usepackage{booktabs}
\usepackage{graphicx}
\usepackage{subfigure}
\usepackage[colorlinks,linkcolor=black,anchorcolor=black,citecolor=black,urlcolor=blue]{hyperref}
\usepackage{amsmath,bm}
\usepackage{booktabs}
\usepackage{mathtools}
\usepackage{amssymb}
\usepackage{caption}
\usepackage{capt-of}
\usepackage{mciteplus}
\usepackage{cite}
\usepackage{mathrsfs}
\usepackage[title,titletoc,toc]{appendix}
\usepackage{xr}
\usepackage{parskip}
\usetikzlibrary{automata,positioning}
\usepackage{wrapfig}
\renewcommand{\part}[1]{\textbf{\large Part \Alph{partCounter}}\stepcounter{partCounter}\\}

\begin{document}

\title{Topological AI forecasting of future dominating viral variants
 } 
\author{Guo-Wei Wei$^{1,2,3}$\footnote{
Corresponding author.		E-mail: weig@msu.edu} \\
$^1$ Department of Mathematics, \\
Michigan State University, MI 48824, USA.\\
$^2$ Department of Electrical and Computer Engineering,\\
Michigan State University, MI 48824, USA. \\
$^3$ Department of Biochemistry and Molecular Biology,\\
Michigan State University, MI 48824, USA. \\
}


\maketitle
 \begin{abstract}
   The understanding of the mechanisms of SARS-CoV-2 evolution and transmission is one of the greatest challenges of our time. By integrating artificial intelligence (AI), viral genomes isolated from patients, tens of thousands of mutational data, biophysics, bioinformatics, and algebraic topology, the SARS-CoV-2 evolution was revealed to be governed by infectivity-based natural selection. Two key mutation sites, L452 and N501 on the viral spike protein receptor-binding domain (RBD), were predicted in summer 2020, long before they occur in prevailing variants Alpha, Beta, Gamma, Delta, Kappa, Theta, Lambda, Mu, and Omicron. Recent studies identified a new mechanism of natural selection: antibody resistance. AI-based forecasting of Omicron’s infectivity, vaccine breakthrough, and antibody resistance was later nearly perfectly confirmed by experiments. The replacement of dominant BA.1 by BA.2 in later March was predicted in early February. On May 1, 2022, persistent Laplacian-based AI projected Omicron BA.4 and BA.5 to become the new dominating COVID-19 variants. This prediction became reality in late June. Topological AI models offer accurate prediction of mutational impacts on the efficacy of monoclonal antibodies (mAbs).  
 \end{abstract}
\begin{figure}
	\centering
	\includegraphics[width=0.5\textwidth]{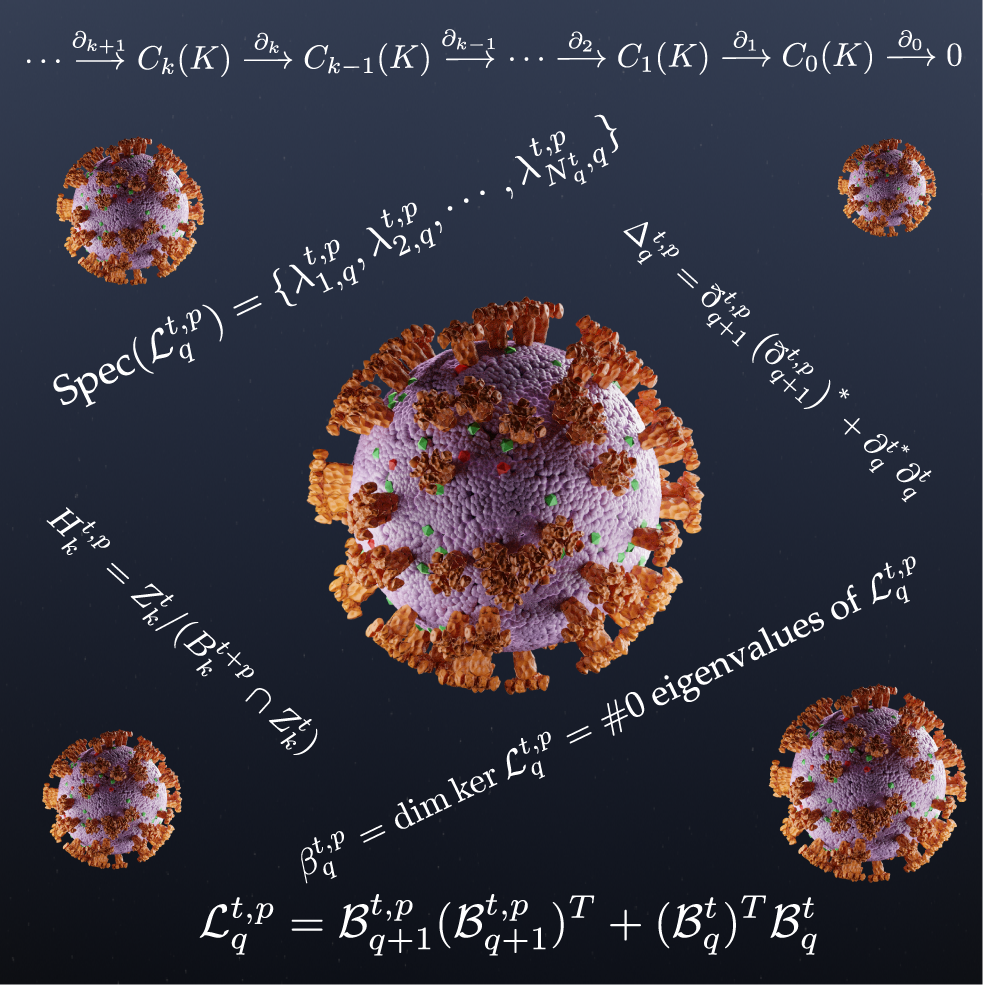}
    \caption{Topology deciphers the virus code.   Image credit: Rui Wang
		}
    \label{fig:TOC}
\end{figure}

SARS-CoV-2 is an extremely sophisticated virus with 29 different proteins. This number includes the spike protein, which enables viral cell entry through its interaction with human angiotensin-converting enzyme 2 (hACE2) at the virus spike receptor-binding domain (RBD). The strength of this interaction is proportional to virus infectivity \cite{walls2020structure}.  Mutations may occur randomly, but natural selection favors RBD mutations that strengthen viral infectivity and evolutionary fitness  \cite{chen2020mutations}. For example, approximately 32 of the 50 mutations of COVID-19’s Omicron variant are located on the spike protein; non-proportionally, 15 are on RBD to optimize viral evolutionary advantages. \cite{chen2022omicron}.  

The spike protein is also the main antigenic target of COVID-19 antibodies that are generated by either infection or vaccination. Spike protein-bound antibodies prevent the SARS-CoV-2 virus from interacting with hACE2 and subsequently block viral cell entry, while antibodies that compete with hACE2 on the spike RBD can directly neutralize the virus. The non-covalent binding between a viral spike protein and an antibody works like a zipper, with the virus spike RBD acting as the zipper’s upper teeth and the antibody serving as the lower teeth. The RBD mutations cause the zipper to misalign or break off, which leads to the (partial) loss of antibody protection and potential reinfection. In some cases, one or several of the vital RBD mutations can significantly enhance RBD-hACE2 binding and dramatically disrupt the binding between the spike RBD and protective antibodies.

Forecasting of the emerging dominant variants helps policymakers plan preventive measures and allows biopharmaceutical companies additional time to develop future vaccines and antibody drugs. However, such forecasting is one of the most challenging scientific tasks of our time. Identifying the mutations that are vital for virus evolution involves striking complexity; a spike protein consists of more than 1,700 amino acid residues, and its dynamical degree of freedom exceeds 5,100. In contrast, the million-dollar Navier-Stokes existence and smoothness problem concerns only three-dimensional dynamics. Additionally, each residue can mutate into one of 19 alternative amino acids with a wide range of chemical, physical, and biological disparities. This possibility creates an astronomically large mutational space that is scaled as  20$^N$,  (where $N$ is the number of involved amino acid residues), thus making full experimental deep mutational screening unfeasible. Moreover, each set of mutations may potentially contribute to a new viral variant. The genotype-phenotype mapping between a variant and its infectivity and/or antibody resistance is highly nonlinear and involves intricate geometric and combinatorial complexities. Innovative strategies are therefore necessary. 
   
Topology offers a solution to this intriguing problem   (see Figure \ref{fig:TOC}).  Traditional topology addresses the invariants of a geometric object under continuous deformation; the homeomorphisms and homotopies may not refer to metrics or coordinates and are too abstract to find use in biological analysis. However, persistent homology—a new branch of algebraic topology that employs multiscale analysis to bridge the gap between complex geometry and abstract topology  \cite{carlsson2009topology,edelsbrunner2008persistent} --- effectively simplifies biomolecular complexity. Persistent homology analyzes biomolecular data in terms of a simplicial complex. The underlying topological space is equipped with filtration to create a family of simplicial complexes, which is a nested sequence of multiscale subsets. But biomolecular systems involve a wide range of interactions ---like covalent bonds, hydrogen bonds, van der Waals, electrostatics, hydrophilicity, hydrophobicity, and so forth --- that the whole molecular persistent homology analysis would miss. Element-specific persistent homology (ESPH) overcomes this obstacle by embedding physical, chemical, and biological information in topological invariants. The power of ESPH was exemplified through its dominating victories in worldwide annual competitions about computer-aided drug design \footnote{\url{https://drugdesigndata.org/about/grand-challenge}}, which is one of the most competitive fields in modern science  \cite{nguyen2019mathematical}. It remains to be seen whether this topological tool can withstand the outburst challenges that are associated with the ongoing COVID-19 pandemic.   
 
Right before the pandemic began, researchers developed an ESPH-based deep learning method that offers state-of-the-art predictions of mutation-induced binding affinity changes of protein-protein interactions --- including antibody-antigen, interactions \cite{wang2020topology}. Scientists integrated this method with genotyping and sequence alignment to create an artificial intelligence (AI) platform, which revealed that SARS-CoV-2 evolution and transmission follows Darwin's natural selection  \cite{chen2020mutations}. The study singled out two vital spike protein residues at positions 452 and 501 that ``have very high chances to mutate into significantly more infectious COVID-19 strains'' long before they occurred in prevailing SARS-CoV-2 variants, such as Alpha, Beta, Gamma, Delta, Epsilon, Theta, Kappa, Lambda, Mu, and Omicron. IDuring this process, ESPH delineates the crucial geometric and biophysical characteristics of mutants in the astronomically large topological space that contributes to virus infectivity, vaccine breakthrough, and antibody resistance.

When reports of the new Omicron variant first emerged in late November 2021, no relevant data was available because experiments had not yet been performed. But within a few days, a topology-based AI platform forecasted Omicron to be nearly three times more infectious than Delta, capable of escaping nearly 90 percent of vaccines, and resistant to essentially all U.S. Food and Drug Administration-approved monoclonal antibodies. Experiments confirmed these predictions in the following weeks  \cite{chen2022omicron}. In early 2022, a new subvariant of Omicron called BA.2 started spreading. On February 11, the same topology-based AI platform forecasted BA.2 as the next dominant variant  \cite{chen2022omicron2}. Six weeks later on March 26, the World Health Organization announced BA.2’s global dominance. 
  
\begin{figure}[h]
	\centering
	\includegraphics[width = 0.8\textwidth]{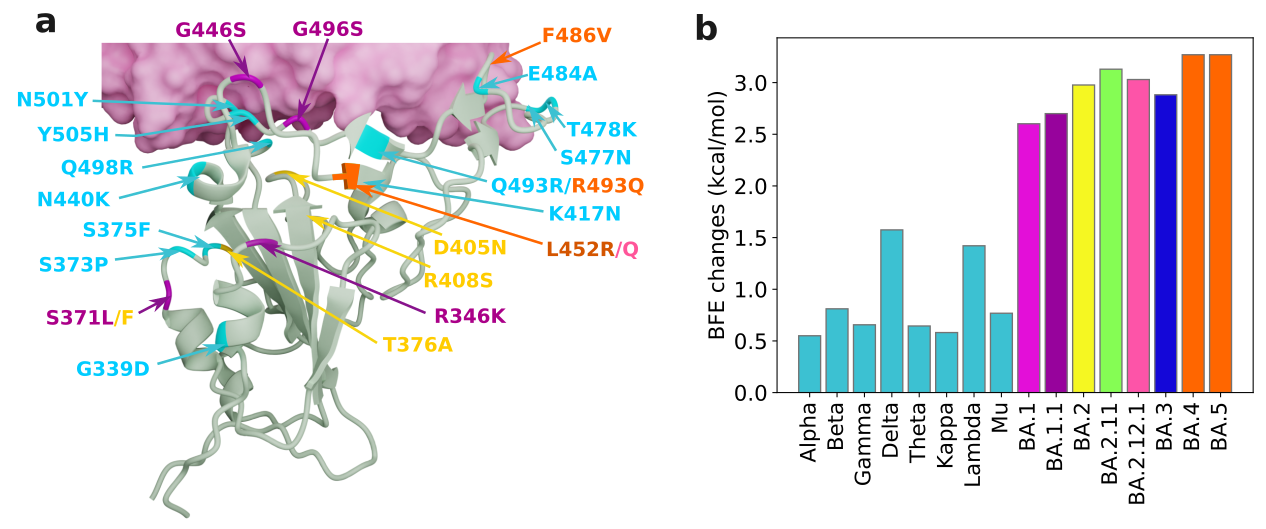}
	\caption{ 
	The receptor-binding domain (RBD) mutations of Omicron subvariants at the RBD-hACE2 interface, as well as their mutation-induced changes in binding free energy (BFE)  \cite{chen2022persistent}. 
{\bf a} RBD mutations of Omicron subvariants at the RBD-hACE2 interface (PDB: 7T9L). 
The shared 12 mutations are shown in cyan. 
BA.1  mutations are plotted with magenta. 
BA.2  mutations are marked in yellow. 
BA.4 and BA.5  mutations are labeled in orange. 
The rest colors can be matched from the right chart.  
{\bf b} A comparison of predicted mutation-induced BFE changes for various SARS-CoV-2 variants and subvariants. According to the Boltzmann distribution, a variant with higher BFE change has an exponential advantage in infectivity. 
Image courtesy of Jiahui Chen. 
	}
	\label{fig:omicron}
\end{figure}
Many Omicron subvariants were circulating around the world by late April, including BA.1, BA.1.1, BA.2, BA.2.11, BA.2.12.1, BA.3, BA.4, and BA.5 (see Figure \ref{fig:omicron}). These subvariants involve in numerous spike protein RBD mutations with very subtle differences, therefore demanding more discriminative mathematical tools.  Although persistent homology is an outstanding tool for the characterization of topological invariants, it is insensitive to the homotopic shape variations in protein-protein interactions that are crucial to viral evolution and transmission. A recent study tackled this challenge with the persistent Laplacian (also known as the persistent spectral graph): a topological Laplacian that is designed to capture both the topological persistence and homotopical shape evolution of data  \cite{wang2020persistent}.   Its harmonic spectra fully recover the topological invariants of persistent homology, while its nonharmonic spectra unveil homotopical shape evolution. On May 1, 2022, persistent Laplacian-based AI projected Omicron BA.4 and BA.5 to become the new dominating COVID-19 variants  \cite{chen2022persistent}. This prediction became reality in late June.  

\section*{Acknowledgment}
This work was supported in part by NIH grants  R01GM126189 and  R01AI164266, NSF grants DMS-2052983,  DMS-1761320, and IIS-1900473,  NASA grant 80NSSC21M0023,  Michigan Economic Development Corporation, MSU Foundation,  Bristol-Myers Squibb 65109, and Pfizer. 

\vspace{1cm}

 \bibliographystyle{abbrv}


\begin{thebibliography}{10}

\bibitem{carlsson2009topology}
G.~Carlsson.
\newblock Topology and data.
\newblock {\em Bulletin of the American Mathematical Society}, 46(2):255--308,
  2009.

\bibitem{chen2022persistent}
J.~Chen, Y.~Qiu, R.~Wang, and G.-W. Wei.
\newblock Persistent Laplacian projected Omicron BA.4 and BA.5 to become new
  dominating variants.
\newblock {\em arXiv preprint arXiv:2205.00532}, 2022.

\bibitem{chen2022omicron}
J.~Chen, R.~Wang, N.~B. Gilby, and G.-W. Wei.
\newblock Omicron variant (B.1.1.529): Infectivity, vaccine breakthrough, and
  antibody resistance.
\newblock {\em J. Chem. Inf. Model.}, 62(2):412--422, 2022.

\bibitem{chen2020mutations}
J.~Chen, R.~Wang, M.~Wang, and G.-W. Wei.
\newblock Mutations strengthened {SARS-CoV-2} infectivity.
\newblock {\em J. Mol. Biol.}, 432(19):5212--5226, 2020.

\bibitem{chen2022omicron2}
J.~Chen and G.-W. Wei.
\newblock Omicron BA.2 (B.1.1.529.2): High potential for becoming the next
  dominant variant.
\newblock {\em The Journal of Physical Chemistry Letters}, 13:3840--3849, 2022.

\bibitem{edelsbrunner2008persistent}
H.~Edelsbrunner and J.~Harer.
\newblock Persistent homology-a survey.
\newblock {\em Contemporary Mathematics}, 453:257--282, 2008.

\bibitem{nguyen2019mathematical}
D.~D. Nguyen, Z.~Cang, K.~Wu, M.~Wang, Y.~Cao, and G.-W. Wei.
\newblock Mathematical deep learning for pose and binding affinity prediction
  and ranking in {D3R Grand Challenges}.
\newblock {\em Journal of Computer-aided Molecular Design}, 33(1):71--82, 2019.

\bibitem{walls2020structure}
A.~C. Walls, Y.-J. Park, M.~A. Tortorici, A.~Wall, A.~T. McGuire, and
  D.~Veesler.
\newblock Structure, function, and antigenicity of the {SARS-CoV-2} spike
  glycoprotein.
\newblock {\em Cell}, 2020.

\bibitem{wang2020topology}
M.~Wang, Z.~Cang, and G.-W. Wei.
\newblock A topology-based network tree for the prediction of protein--protein
  binding affinity changes following mutation.
\newblock {\em Nature Machine Intelligence}, 2(2):116--123, 2020.

\bibitem{wang2020persistent}
R.~Wang, D.~D. Nguyen, and G.-W. Wei.
\newblock Persistent spectral graph.
\newblock {\em International Journal for Numerical Methods in Biomedical
  Engineering}, 36(9):e3376, 2020.

\end{thebibliography}

\end{document}